\documentclass[aps,twocolumn,preprintnumbers,nofootinbib,superscriptaddress]{revtex4}
\usepackage{amsmath} \usepackage{graphicx} \usepackage{amsfonts}
\usepackage{array} \usepackage{amsthm} \usepackage{bm}
\usepackage{palatino} \usepackage{mathpazo} 
\usepackage{supertabular}

\usepackage[breaklinks]{hyperref}
\usepackage{color}

\usepackage{epstopdf}

\newcommand{\be}{\begin{equation}}
\newcommand{\ee}{\end{equation}}
\newcommand{\ba}{\begin{eqnarray}}
\newcommand{\ea}{\end{eqnarray}}
\newcommand{\bal}{\begin{align}}
\newcommand{\eal}{\end{align}}

\newcommand{\dd}{{\rm d}}

\newcommand{\La}{\Lambda}
\newcommand{\bt}{\beta}

\newcommand{\ga}{\gamma}

\newcommand{\Om}{\Omega}

\newcommand{\bw}{\begin{widetext}}
\newcommand{\ew}{\end{widetext}}

\begin{document}

\title{Cyclic and heteroclinic flows near general static spherically symmetric black holes: Semi-cyclic flows -- Addendum and corrigendum}

\author{Mustapha Azreg-A\"{\i}nou}
\affiliation{Engineering Faculty, Ba\c{s}kent University, Ba\u{g}l\i ca Campus, Ankara, Turkey}


\begin{abstract}
We present new accretion solutions of a polytropic perfect fluid onto an f(R)-gravity de Sitter-like black hole. We consider two f(R)-gravity models and obtain finite-period cyclic flows oscillating between the event and cosmological horizons as well as semi-cyclic critical flows executing a two-way motion from and back to the same horizon. Besides the generalizations and new solutions presented in this work, a corrigendum to Eur. Phys. J. C (2016) 76:280 is provided.
\end{abstract}

\maketitle



\section{Accretion of perfect fluids onto static spherically symmetric black holes}

This work is based on our previous paper~\cite{aafj} where we set the general dynamical-system formalism for accretion of perfect fluids onto static spherically symmetric black holes. We keep using the same notation for the thermodynamic functions of the fluid and the Hamiltonian $\mathcal{H}$. So, $n$, $h$, $e$, $s$, $p$, $T$, and $u^{\mu}$ are the baryon number density, specific enthalpy (enthalpy per particle), energy density, specific entropy, pressure, temperature, and four-velocity vector, respectively. In a locally inertial frame, the three-dimensional speed of sound $a$ is given by $a^2=(\partial p/\partial e)_s$. When the entropy $s$ is constant, which is the case for accretion of perfect fluids onto static spherically symmetric black holes, this reduces to $a^2=dp/de$.

\subsection{Consequences of the conservation laws - Thermodynamics}

The aim of this short work is two-fold: (1) Generalize the dynamical-system formalism for accretion of perfect fluids to metrics of the form
\begin{equation}\label{1}
\dd s^{2}= -A(r)\dd t^2+\frac{\dd r^{2}}{B(r)}+C(r)(\dd \theta^2+\sin^2\theta \dd \phi^2).
\end{equation}
This will allow us to generalize the properties of the accreting fluid intended for future use. (2) Obtain new interesting solutions not discussed so far in the scientistic literature.

In~\eqref{1}, ($A,\,B,\,C$) are any functions of the radial coordinate $r$ assumed to be well-defined and positive-definite in the regions where the Killing vector $\xi^{\mu}=(1,0,0,0)$ is timelike and their ratio $D\equiv A/B$ is positive-definite on the horizons too. For the metric~\eqref{1} to describe a black hole solution, the equations $B(r)=0$ and $A(r)=0$ should have the same set of solutions with same multiplicities. In our applications we restrict ourselves to the cases where the global structure of the spacetime is well-determined by ($A,\,B,\,C$). To keep the analysis general, we do not assume asymptotic flatness of the metric as we intend to apply it to the de Sitter and anti-de Sitter-like black holes.

This metric form generalizes the one used in Refs.~\cite{aafj,aafs,CS}; In Ref.~\cite{aafj} we restricted ourselves to the case $A=B=f(r)$ and $C=r^2$. It is easy to show that equations~(5), (7), (11), (23), (24), (25) of Ref.~\cite{aafj} generalize respectively to
\begin{align}
\label{2a}&u_t=\pm \sqrt{A+Du^2}, &\sqrt{D}~Cnu=C_1\neq 0,\\
\label{2b}&h\sqrt{A+Du^2}=C_2, &v^2=\frac{Du^2}{A+Du^2},\\
\label{2c}&u^2=\frac{Bv^2}{1-v^2}, &\frac{AC^2n^2v^2}{1-v^2}=C_1^2,
\end{align}
where ($C_1,C_2$) are constants of motion, $u\equiv u^r$, and $-1<v<1$ is the three-velocity of a fluid element as measured by a locally static observer. The second line in~\eqref{2c} expresses the law of particle conservation, $\nabla_{\mu}(n u^{\mu})=0$, and $C_2$ is the constant of motion $hu_{\mu}\xi^{\mu}$ [$\xi^{\mu}=(1,0,0,0)$ is a timelike Killing vector]. This constant is the inertial-equivalent generalization of the energy conservation equation $mu_{\mu}\xi^{\mu}$~\cite{Rezzolla}.

The constant $C_1^2$ in~\eqref{2c} can be written as $A_0C_0^2n_0^2v_0^2/(1-v_0^2)$ where ``0" denotes any reference point ($r_0,\,v_0$) from the phase portrait; this could be a CP, if there is any, spatial infinity ($r_{\infty},\,v_{\infty}$), or any other reference point. We can thus write
\begin{equation}\label{6}
\frac{n^2}{n_0^2}=\frac{A_0C_0^2v_0^2}{1-v_0^2}~\frac{1-v^2}{AC^2v^2}=C_1^2~\frac{1-v^2}{AC^2v^2}.
\end{equation}
The equation of state (EoS) of the fluid may be given in the form $e=F(n)$ or equivalently in the form $p=G(n)$~\cite{aafj}. It has been shown~\cite{aafj} that ($F,\,G$), the specific enthalpy, and the three-dimensional speed of sound satisfy
\begin{align}
\label{5a}&nF'(n)-F(n)=G(n),\\
\label{5b}&h=F'(n),\\
\label{5c}&a^2=n(\ln F')',
\end{align}
where the prime denotes derivative with respect to $n$.

Equations~\eqref{6} and~\eqref{5b} imply that the specific enthalpy $h$ depends explicitly on ($A,\,C,\,v$) only. There is no explicit dependence on $B$. As we shall see in the next subsection, this implies that the Hamiltonian $\mathcal{H}$, which defines the dynamical system, will also depend explicitly on ($A,\,C,\,v$) only. This fact will have consequences on the location of the critical points (CP's). If the fluid had further properties, say, being isothermal or polytropic, $h$ and $\mathcal{H}$ can be expressed explicitly in terms of ($A,C$), as this is done in the following section.

Horizons $r_h$ are defined by $A(r_h)=0$ and $B(r_h)=0$ or simply by $B(r_h)=0$ since the equations $A(r_h)=0$ and $B(r_h)=0$ are assumed to have the same set of solutions with same multiplicities. The zeros of $B(r_h)=0$ determine the regions in three-space where the fluid flow takes place: These are the regions where $\xi^{\mu}=(1,0,0,0)$ is timelike.

Note that the case $C_1\equiv 0$, corresponding either to $n=0$~\eqref{2a} (no fluid) or to $v=0$~\eqref{2c} and any $n$ (no flow), is not interesting. So, we assume $C_1\neq 0$. As the fluid approaches, or emanates from, any horizon ($r\to r_h$), $A$ approaches 0. Three cases emerge from~\eqref{6}:
\begin{enumerate}
  \item $v\to \pm 1^{\mp}$ and $n$ may converge or diverge there ($r\to r_h$);
  \item $v\to 0$ and $n$ diverges there. The fluid cumulates near the horizon resulting in a divergent pressure which repulses the fluid backwards~\cite{aafj};
  \item $|v|$ assumes any value between 0 and 1 there. This yields a divergent $n$ as $r\to r_h$. Since $0<|v|<1$ there is no reason that the fluid cumulates near the horizon: The flow continues until all fluid particles have crossed the horizon. We rule out this case for it is not physical. This conclusion, due to the law of particle conservation, is general and it does not depend on the fluid characteristics.
\end{enumerate}
Since a non-perfect simple fluid (containing a single particle species) also obeys the law of particle conservation~\eqref{2c}, these conclusions remain valid for real fluids too.

\subsection{Dynamical system - Critical points}

If the fluid had a uniform pressure, that is, if the fluid were not subject to acceleration, the specific enthalpy $h$ reduces to the particle mass $m$ and the first equation in~\eqref{2b} reduces to $mu_{\mu}\xi^{\mu}=C_2$ along the fluidlines. This is the well know energy conservation law which stems from the fact that the fluid flow is in this case geodesic. Now, if the pressure throughout the fluid is not uniform, acceleration develops through the fluid and the fluid flow becomes non-geodesic; the energy conservation equation $mu_{\mu}\xi^{\mu}=cst$, which is no longer valid, generalizes to its relativistic equivalent~\cite{Rezzolla} $hu_{\mu}\xi^{\mu}=C_2$ as expressed in the first equation in~\eqref{2b}.

Let the Hamiltonian $\mathcal{H}$ of the dynamical system be proportional to $C_2^2$~\eqref{2c}, which is a constant of motion. Substituting the first equation in~\eqref{2c} into the first equation in~\eqref{2b} yields
\begin{equation}\label{3}
\mathcal{H}(r,v)=\frac{h(r,v)^2A(r)}{1-v^2}.
\end{equation}

With $\mathcal{H}$ given by~\eqref{3}, the dynamical system reads
\begin{equation}\label{4}
\dot{r}=\mathcal{H}_{,v}\,,  \quad\quad \dot{v}=-\mathcal{H}_{,r}.
\end{equation}
(here the dot denotes the $\bar{t}$ derivative where $\bar{t}$ is the time variable of the Hamiltonian dynamical system). In~\eqref{4} it is understood that $r$ is kept constant when performing the partial differentiation with respect to $v$ in $\mathcal{H}_{,v}$ and that $v$ is kept constant when performing the partial differentiation with respect to $r$ in $\mathcal{H}_{,r}$. The critical points (CPs) of the dynamical system are the points ($r_c,v_c$) where the rhs's in~\eqref{4} are zero. To take advantage of the calculations made in~\cite{aafj} we introduce the radial coordinate $\rho$ and the notation $f$ defined by
\begin{equation}\label{d1}
\rho^2(r)\equiv C(r),\quad\quad f(\rho)\equiv A(r).
\end{equation}
The Hamiltonian takes the form
\begin{equation}
\mathcal{H}(\rho,v)=\frac{h(\rho,v)^2f(\rho)}{1-v^2}.
\end{equation}
The derivative $\mathcal{H}(\rho,v)_{,\rho}$ has been evaluated in Ref.~\cite{aafj} by [Eq.~(40) of Ref.~\cite{aafj}]:
\begin{equation}\label{d2}
    \mathcal{H}_{,\rho}=\frac{h^2}{1-v^2}\Big[\frac{\dd f}{\dd \rho}-2a^2f~\frac{\dd \ln(\sqrt{f}\rho^2)}{\dd \rho}\Big].
\end{equation}
Using
\begin{multline*}
\mathcal{H}_{,r}=\frac{\dd \rho}{\dd r}~\mathcal{H}_{,\rho}=\frac{h^2}{1-v^2}\Big[\frac{\dd f}{\dd r}-2a^2f~\frac{\dd \ln(\sqrt{f}\rho^2)}{\dd r}\Big]\\
=\frac{h^2}{1-v^2}\Big[\frac{\dd A}{\dd r}-2a^2A~\frac{\dd \ln(\sqrt{A}C)}{\dd r}\Big],
\end{multline*}
we obtain
\begin{align}
\label{d3}&\dot{r}=\frac{2h^2A}{v(1-v^2)^2}~(v^2-a^2),\\
\label{d4}&\dot{v}=-\frac{h^2}{1-v^2}\Big[\frac{\dd A}{\dd r}-2a^2A~\frac{\dd \ln(\sqrt{A}C)}{\dd r}\Big].
\end{align}
Introducing the notation $g_c=g(r)|_{r=r_c}$ and $g_{c,r_c}=g_{,r}|_{r=r_c}$ where $g$ is any function of $r$, the following equations provide a set of CPs that are solutions to $\dot{r}=0$ and $\dot{v}=0$:
\begin{multline}\label{cp}
v_c^2=a_c^2\quad\text{ and }\\ a_c^2=\frac{C_cA_{c,r_c}}{C_cA_{c,r_c}+2AC_{c,r_c}}=\frac{C^2A_{,r}}{(C^2A)_{,r}}\Big|_{r=r_c},
\end{multline}
where $a_c$ is the three-dimensional speed of sound evaluated at the CP. The first equation states that at a CP the three-velocity of the fluid equals the speed of sound. The second equation determines $r_c$ once the EoS, $e=F(n)$ or $p=G(n)$~\eqref{5a}, is known.

From the set of equations~\eqref{cp} we see that the metric function $B$ does not enter explicitly in the determination of the CP's; however, it does that implicitly via the successive dependence of $h$ on $n$, of $n$ on $v$, and of $v$ on $D$.

Other sets of CPs, solutions to $\dot{r}=0$ and $\dot{v}=0$, may exist too. For instance, we may have (1) $A_c=0$ and $A_{c,r_c}=0$ which, by~\eqref{d3} and~\eqref{d4}, yield $\dot{r}=0$ and $\dot{v}=0$ without having to impose the constraint $v_c^2=a_c^2$ at the CP. This corresponds to a double-root horizon of an extremal black hole. When this is the case, the accretion becomes transonic well before the fluid reaches the CP, which is the horizon itself (recall that the three-velocity $v$, as the fluid approaches the horizon, tends to $-1$). However, extremal black holes are unstable and whatever accretes onto the hole modifies its mass making it non-extremal so that $A_{c,r_c}=0$ no longer holds. We may also have (2) $h=0$ at some point where $\xi^{\mu}=(1,0,0,0)$ is timelike, which may hold only for non-ordinary (dark, phantom, or else) accreting matter.

\section{Specific perfect fluids}

\subsection{Hamiltonian system for test isothermal perfect fluids}
The isothermal EoS is of the form $p=ke=kF(n)$ with $G(n)=kF(n)$ where $k$, the so-called state parameter, obeys the constraints $0<k\leq1$. The differential equation~\eqref{5a} reads
\begin{equation}\label{i1}
nF'(n)-F(n)=kF(n),
\end{equation}
yielding
\begin{equation}\label{i2}
    e=F=\frac{e_0}{n_0^{k+1}}\,n^{k+1},\quad h=\frac{(k+1)e_0}{n_0}\Big(\frac{n^2}{n_0^2}\Big)^{k/2},
\end{equation}
where we have used~\eqref{5b}. Using this and~\eqref{6} in~\eqref{3} we obtain
\begin{equation}\label{i3}
\mathcal{H}(r,v)=\frac{A(r)^{1-k}}{C(r)^{2 k} v^{2 k}(1-v^2)^{1-k}},
\end{equation}
where all the constant factors have been absorbed into the  redefinition of the time $\bar{t}$ and the Hamiltonian $\mathcal{H}$.

\subsection{Hamiltonian system for test polytropic perfect fluids}
The polytropic equation of state is
\begin{equation}\label{p1}
p=G(n)=\mathcal{K}n^{\gamma},
\end{equation}
where $\mathcal{K}$ and $\gamma>1$ are constants. Inserting~\eqref{p1} into the differential equation~\eqref{5a}, it is easy to determine the specific enthalpy by integration~\cite{aafj}
\begin{equation}\label{p2}
  h=m+\frac{\mathcal{K}\ga n^{\ga-1}}{\ga-1},
\end{equation}
where we have introduced the baryonic mass $m$. Introducing the constant
\begin{equation}\label{p3}
Y\equiv \frac{\mathcal{K}\ga (C_1n_0)^{\ga-1}}{m(\ga-1)}=\text{ const.}>0,
\end{equation}
then using~\eqref{6}, $h$ takes the form
\begin{equation}\label{p4}
    h=m\Big[1+Y\Big(\frac{1-v^2}{AC^2v^2}\Big)^{(\ga-1)/2}\Big].
\end{equation}
Finally, the Hamiltonian~\eqref{3} reduces to
\begin{equation}\label{p5}
   \mathcal{H}=\frac{A}{1-v^2}~\Big[1+Y\Big(\frac{1-v^2}{AC^2v^2}\Big)^{(\ga-1)/2}\Big]^2,
\end{equation}
where $m^2$ has been absorbed into a re-definition of ($\bar{t},\mathcal{H}$).

The three-dimensional speed of sound is obtained from~\eqref{5c}
\begin{equation}\label{p6}
   a^2=\frac{(\ga -1)X}{m(\ga -1)+X}\qquad (X\equiv \mathcal{K}\ga n^{\ga -1}),
\end{equation}
Since $\ga >1$, this implies $a^2<\ga -1$ and, particularly, $v_c^2<\ga -1$ if there are CPs to the Hamiltonian system. Equation~\eqref{p6} bears a striking similarity with Eq.~(2.249) on page 119 of Ref.~\cite{Rezzolla}.

\subsection*{Corrigendum}
Using the second equation in~\eqref{2c} and~\eqref{p3}, we rewrite $X$~\eqref{p6} as
\begin{equation}
    X=m(\ga-1)Y\Big(\frac{1-v^2}{AC^2v^2}\Big)^{(\ga-1)/2}.
\end{equation}
Substituting into~\eqref{p6}, we arrive at
\begin{equation}\label{p7}
a^2=Y(\ga -1-a^2)\Big(\frac{1-v^2}{AC^2v^2}\Big)^{(\ga-1)/2}.
\end{equation}
This equation along with the second line in~\eqref{cp} take the following expressions at the CPs
\begin{align}
\label{p8}&v_c^2=Y(\ga -1-v_c^2)\Big(\frac{1-v_c^2}{A_cC_c^2v_c^2}\Big)^{(\ga-1)/2},\\
\label{p9}&v_c^2=\frac{r_cA_{c,r_c}}{r_cA_{c,r_c}+4A_c}.
\end{align}
For a given value of the positive constant $Y$, the resolution of this system of equations in ($r_c,v_c$) provides all the CPs, if there are any.

In Ref.~\cite{aafj} we worked with $A=B=f$ and $C=r^2$ reducing~\eqref{p8} to
\begin{equation}\label{p10}
v_c^2=Y(\ga -1-v_c^2)\Big(\frac{1-v_c^2}{r_c^4f_cv_c^2}\Big)^{(\ga-1)/2},
\end{equation}
which is the correct expression of Eq.~(112) of Ref.~\cite{aafj}. In both equations~(111) and (112) of Sec.~VI of Ref.~\cite{aafj}, one should replace the constant factor
\begin{equation*}
    \frac{n_c}{Y}~\Big(\frac{r_c^5f_{c,r_c}}{4}\Big)^{1/2}
\end{equation*}
by 1. The presence of this extra factor did not affect the results, solutions, and conclusions made in Ref.~\cite{aafj}; however, some new interesting solutions have been missed in its Sec.~VI on polytropic fluids. Besides the results and conclusions we have discussed so far in the two first sections of this work, we aim (1) to re-derive the same solutions derived in Sec.~VI of Ref.~\cite{aafj}, using the correct expression~\eqref{p10}, and (2) to construct new solutions.

\section{Accretion of polytropic test fluids\label{secPTS}}

\subsection{f(R)-gravity model of Ref.~\cite{q4}}
In Ref.~\cite{aafj}, we considered three models of f(R) gravity~\cite{q4,m1,m2}. For the model of Ref.~\cite{q4} the black hole solution is of the form $A=B=f$ and $C=r^2$ with
\begin{equation}\label{a1}
f\equiv 1-\frac{2M}{r}+\beta r-\frac{\Lambda r^{2}}{3}.
\end{equation}
Following the notation of Ref.~\cite{aafj}, we employ in this work the symbol ``$f$" for the metric component, $-g_{tt}$, and the symbol ``f" for the function f(R) defining the f(R)-gravity model.

\begin{figure*}[!htb]
\centering
\includegraphics[width=0.43\textwidth]{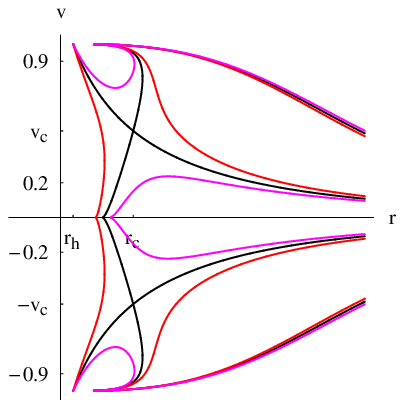} \includegraphics[width=0.43\textwidth]{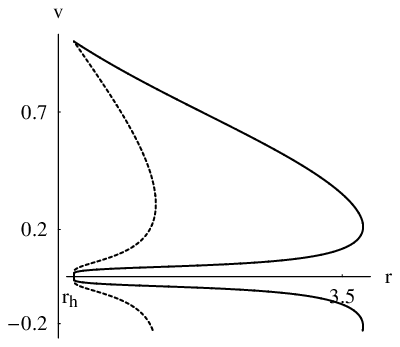}\\
\caption{{\footnotesize Left panel is a contour plot of $\mathcal{H}$~\eqref{p5} for an anti-de Sitter-like $\text{f}(R)$ black hole~\eqref{a1} with $M=1$, $\bt =0.05$, $\La =-0.04$, $\ga=1/2$, $Y=-1/8$. The parameters are $r_h\simeq 1.76955$, $r_c\simeq 3.68101$, $v_c\simeq 0.498889$. Black plot: the solution curve through the CPs $(r_{c},v_c)$ and $(r_{c},-v_c)$ for which $\mathcal{H}=\mathcal{H}_c\simeq 0.487469$. Red plot: the solution curve for which $\mathcal{H}=\mathcal{H}_c - 0.09$. Magenta plot: the solution curve for which $\mathcal{H}=\mathcal{H}_c + 0.09$. Right panel is a contour plot of $\mathcal{H}$~\eqref{p5} for an anti-de Sitter-like $\text{f}(R)$ black hole~\eqref{a1} with $M=1$, $\bt =0.05$, $\La =-0.04$, $\ga =5.5/3$, $Y=1/8$. The horizon is at $r_h\simeq 1.76955$ and there are no CPs. Continuous black plot: the solution curve corresponding to $\mathcal{H}=0.94447$. Dashed black plot: the solution curve corresponding to $\mathcal{H}=0.443809$. For the clarity of the plot, we have partially removed the branches $v<0$.}}\label{Fig1}
\end{figure*}

The solutions shown in Fig.~5 of Ref.~\cite{aafj}, which depict the accretion of a polytropic perfect fluid onto an anti-de Sitter-like $\text{f}(R)$ black hole~\eqref{a1}, are re-derived using the same values of the parameters: $M=1$, $\bt =0.05$, $\La =-0.04$, $\ga=1/2$, and $Y=-1/8$. The re-derived solutions using the correct expressions~\eqref{p8} and~\eqref{p9} are plotted in Fig.~\ref{Fig1} of this work.
\begin{figure*}[!htb]
\centering
\includegraphics[width=0.33\textwidth]{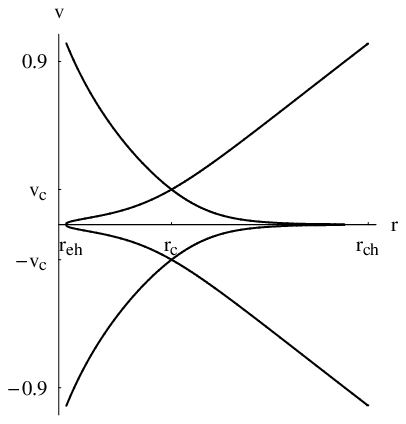} \includegraphics[width=0.33\textwidth]{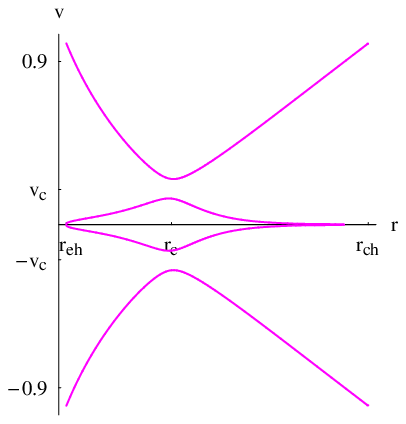} \includegraphics[width=0.33\textwidth]{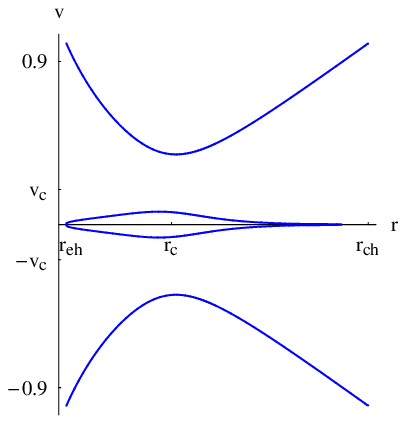}\\
\caption{{\footnotesize Contour plot of $\mathcal{H}$~\eqref{p5} for a de Sitter-like $\text{f}(R)$ black hole~\eqref{a1} with $M=1$, $\bt =0.05$, $\La =0.04$, $\ga =1.7$, $Y=1/8$. The parameters are $r_{eh}\simeq 1.91048$, $r_{ch}\simeq 9.8282$, $r_{c}\simeq 4.66942$, $v_{c}\simeq 0.19387$. Black plot: the solution curve through the CPs $(r_{c},v_{c})$ and $(r_{c},-v_{c})$ for which $\mathcal{H}=\mathcal{H}_{c}\simeq 0.59691$. This solution is new and it was not discovered in Ref.~\cite{aafj}. Magenta plot: the solution curve corresponding to $\mathcal{H}=\mathcal{H}_{c}+ 0.005$. Blue plot: the solution curve corresponding to $\mathcal{H}=\mathcal{H}_{c}+ 0.05$.  The solutions depicted by the magenta and blue plots are not new and were discovered in Ref.~\cite{aafj}.}}\label{Fig2}
\end{figure*}
The first thing we note is that the re-derived solution corresponding to $\ga =5.5/3$, right panel of Fig.~\ref{Fig1}, has no CPs. Apart from this, the re-derived solutions have the same characteristics as those shown in Fig.~5 of Ref.~\cite{aafj}.

\begin{figure*}[!htb]
\centering
\includegraphics[width=0.43\textwidth]{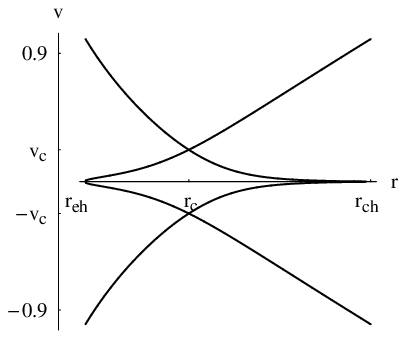} \includegraphics[width=0.43\textwidth]{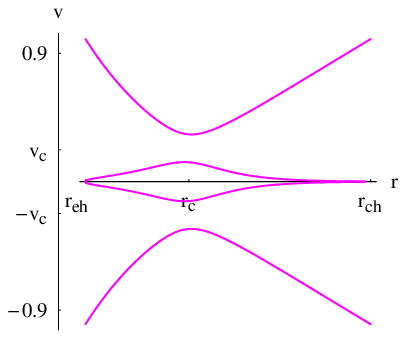}\\
\caption{{\footnotesize Contour plot of $\mathcal{H}$~\eqref{p5} for a de Sitter-like $\text{f}(R)$ black hole with $\text{f}(R)$ given by Hu-Sawicki formula~\eqref{hs5}. We worked with $M=1$, $Q =0.01$, $R_0 =0.16$, $\ga =1.7$, $Y=1/8$, $q_1=41$, $q_2=19$, and $c_1$ and $c_2$ are given by~\eqref{hs8} taking the lower sign corresponding to the physical solution $|\text{f}\,'(R_0)|\ll 1$. Left plot: Critical flow corresponding to $\mathcal{H}=\mathcal{H}_{c}\simeq 0.35067$. Here $\text{f}\,'(R_0) \simeq 0.0372803$, $r_{eh}\simeq 2.12854$, $r_{ch}\simeq 7.3975$, $r_{c}\simeq 4.03815$, $v_{c}\simeq 0.223489$. This solution is new and it was not discovered in Ref.~\cite{aafj}. Right plot: Corresponds to $\mathcal{H}=\mathcal{H}_{c}+0.01$. This solution is not new and was discovered in Ref.~\cite{aafj}.}}\label{Fig3}
\end{figure*}

The solutions depicting the accretion of a polytropic perfect fluid onto a de Sitter-like $\text{f}(R)$ black hole~\eqref{a1}, which are shown in Fig.~\ref{Fig2} of this work, have been constructed using the same values of the parameters used in Fig.~6 of Ref.~\cite{aafj}: $M=1$, $\bt =0.05$, $\La =0.04$, $\ga =1.7$, and $Y=1/8$. The magenta and blue solutions were discovered in Ref.~\cite{aafj}. The new solutions are the semi-cyclic critical black plots of Fig.~\ref{Fig2}. The first semi-cyclic solution represents a supersonic accretion from the cosmological horizon, where the initial three-velocity is almost $-1$, then it becomes subsonic passing the CP, and it vanishes on the event horizon. The accretion is followed by a flowout back to the cosmological horizon reversing all the details. The second semi-cyclic solution is a flowout from the event horizon with an initial three-velocity in the vicinity of $1$, which decreases gradually until it is sonic at the CP then zero at the cosmological horizon. This flowout is then followed by an accretion back to the event horizon.

Notice that on the horizons, $r_h=r_{eh}$ (event horizon) or $r_h=r_{ch}$ (cosmological horizon), the pressure of the fluid diverges as~\cite{aafj}
\begin{equation}\label{a2}
    p\propto |r-r_h|^{\frac{-\ga}{2(\ga-1)}}\qquad (1<\ga<2).
\end{equation}
if there $v=0$. This explains why the fluid, once it reaches any horizon with vanishing three-velocity, it is repulsed backward under the effect of its own pressure.

As the value of the Hamiltonian exceeds the critical value $\mathcal{H}_{c}\equiv \mathcal{H}(r_c,v_c)$, cyclic flows between the two horizons form. These flows are sandwiched by two separate branches corresponding to supersonic accretion and flowout, as depicted by the magenta and blue plots of Fig.~\ref{Fig2}. The separation between these supersonic branches increases with the value of the Hamiltonian resulting in faster accretion and flowout while the cyclic flow tends to become more and more nonrelativistic.

\subsection{f(R)-gravity model of Ref.~\cite{m1}}
For the f(R) model of Ref.~\cite{m1} we constructed the constant-curvature black hole solution~\cite{aafj}
\begin{equation}\label{hs4}
    f(r)=1-\frac{2M}{r}+\frac{Q^2}{[1+\text{f}\,'(R_0)]r^2}-\frac{R_0}{12}~r^2,
\end{equation}
where
\begin{equation}\label{hs5}
\text{f}(R)=-\mathcal{M}^2\frac{c_1(R/\mathcal{M}^2)^n}{c_2(R/\mathcal{M}^2)^n+1}.
\end{equation}
Here $n>0$, ($c_1,c_2$) are proportional constants~\cite{m1}
\begin{equation}\label{hs6}
\frac{c_1}{c_2}\equiv q_2\approx 6~\frac{\Om_{\La}}{\Om_{m}}=6~\frac{0.76}{0.24}=19,
\end{equation}
and the mass scale
\begin{equation*}
\mathcal{M}^2=(8315\text{Mpc})^{-2}\Big(\frac{\Om_{m}h^2}{0.13}\Big).
\end{equation*}
At the present epoch~\cite{m1}
\begin{equation}\label{hs7}
\frac{R_0}{\mathcal{M}^2}\equiv q_1\approx\frac{12}{\Om_{m}}-9=41.
\end{equation}

Taking $n=2$, we found~\cite{aafj}
\begin{equation}\label{hs8}
c_1=q_2c_2,\qquad c_2=-\frac{1}{q_1^{3/2}(\sqrt{q_1}\pm\sqrt{2q_2})}.
\end{equation}
The physical solution corresponds to $|\text{f}\,'(R_0)|\ll 1$. Using the correct equation~\eqref{p10} and Eq.~\eqref{p9}, which takes the form
\begin{equation}\label{hs10}
\hspace{-2mm}v_c^2=\frac{(1+\text{f}\,'(R_0)) (R_0 r_c^3-12 M) r_c+12 Q^2}{3 [(1+\text{f}\,'(R_0)) (R_0 r_c^3-8 r_c+12 M) r_c-4 Q^2]},
\end{equation}
we construct the new solutions, shown in Fig.~\ref{Fig3}, using the same values of the parameters used in Fig.~7 of Ref.~\cite{aafj}: $M=1$, $Q =0.01$, $R_0 =0.16$, $\ga =1.7$, $Y=1/8$, $q_1=41$, $q_2=19$, and $c_1$ and $c_2$ are given by~\eqref{hs8} taking the lower sign corresponding to the physical solution $|\text{f}\,'(R_0)|\ll 1$. These new solutions have the same characteristics of those depicted in Fig.~\ref{Fig2}. Only the semi-cyclic critical black plots represent new solutions: The solution depicted by the magenta plot of Fig.~\ref{Fig3} is not new and was discovered in Ref.~\cite{aafj}.

\section{Conclusion}
In this addendum we have first generalized the dynamical-system procedure describing the accretion/flowout of perfect fluids to all black holes endowed with spherical symmetry. This is needed for many future investigations~\cite{aajr}. In our dynamical-system procedure we took the radial coordinate and the three-velocity as dynamical variables of the Hamiltonian, which is proportional to the square of the constant of motion $hu_{\mu}\xi^{\mu}$ [$\xi^{\mu}=(1,0,0,0)$ is a timelike Killing vector]. This constant is the relativisitc equivalent generalization of the energy conservation equation $mu_{\mu}\xi^{\mu}$~\cite{Rezzolla}.

We have shown that the de Sitter-like black holes, of different f(R)-gravity models, present cyclic non-critical flows and semi-cyclic critical flows all characterized by a vanishing three-velocity on either horizon or a luminal three-velocity there: no situations where the fluid reaches, or emanates from, either horizon with intermediate three-velocity occur. This is due to the law of particle conservation and it is not related to the nature of the fluid. This conclusion remains valid for real fluids too as they approach any horizon from within a region where $\xi^{\mu}=(1,0,0,0)$ is timelike.

\end{document}